\documentclass[superscriptaddress,prl,twocolumn,showpacs,preprintnumbers,amsmath,amssymb]{revtex4}

\usepackage{graphicx}
\usepackage{url}

\begin{document}

\title{Breaking arches with vibrations: the role of defects}

\author{Celia Lozano}
\affiliation{Departamento de F\'{\i}sica, Facultad de Ciencias,
Universidad de Navarra, 31080 Pamplona, Spain.}
\author{Geoffroy Lumay}
\affiliation{GRASP, Institut de Physique, Bat. B5a Sart-Tilman, Universit\'e de Li\`{e}ge, B-4000 Li\`{e}ge, Belgium}
\author{Iker Zuriguel}
\affiliation{Departamento de F\'{\i}sica, Facultad de Ciencias,
Universidad de Navarra, 31080 Pamplona, Spain.}
\author{Angel Garcimart\'{\i}n}\email{angel@fisica.unav.es}
\affiliation{Departamento de F\'{\i}sica, Facultad de Ciencias,
Universidad de Navarra, 31080 Pamplona, Spain.}

\date{\today}

\begin{abstract}
We present experimental results about the stability of arches against external vibrations. Two dimensional strings of mutually stabilizing grains are geometrically analyzed and subsequently submitted to a periodic forcing at fixed frequency and increasing amplitude. The main factor that determines the granular arch resistance against vibrations is the maximum angle among those formed between any particle of the arch and its two neighbors: the higher the maximum angle is, the easier to break the arch. Based in an analysis of the forces, a simple explanation is given for this dependence. From this, interesting information can be extracted about the expected magnitudes of normal forces and friction coefficients of the particles conforming the arches.
\end{abstract}

\pacs{45.70.Mg}

\maketitle

The formation of arches is a common feature whenever large assemblies of solids move collectively. Force chains that propagate along a string of particles are able to stop the movement, in such a way that the ensemble can resist an external pressure as a solid does. Arches, defined as arrangements of mutually stabilizing sets of particles capable of withstanding external loads \cite{Mehta,Pugnaloni}, are the key ingredient for attaining a solid-like structure. The likeness of this can happen, for instance, in traffic jams \cite{Schadschneider,Helbing1}, avalanches of crowds in panic \cite{Helbing2} and colloidal systems \cite{Jenkins}. The notion that a common description can be given for all those various systems was put forward by Cates \emph{et al.} \cite{Cates}, who called them \textit{fragile matter}. They pointed out that arches can be shattered by the exertion of incompatible stresses, meaning forces that act in directions differing from that of the load supported by the arches.

\begin{figure}
\includegraphics[width=0.9\columnwidth]{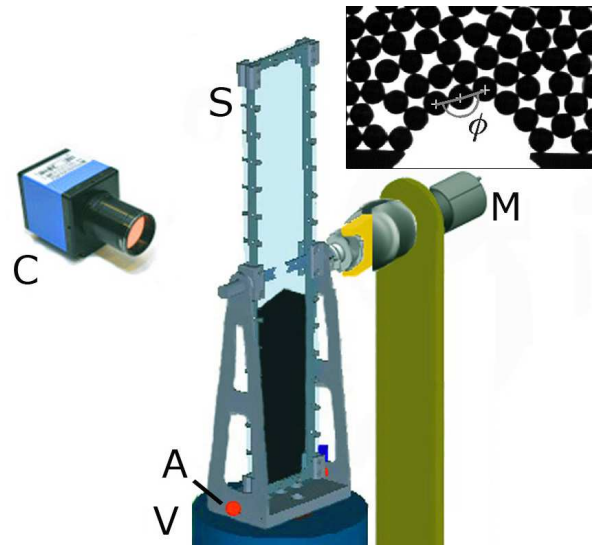}
\caption{(Color online) Sketch of the experimental set-up. C:
video camera; A: accelerometer; V: electromagnetic shaker; M:
motor; S: silo. \emph{Inset}: A photograph of an arch,
showing the angle $\phi$ for one bead.} \label{fig:sistexp}
\end{figure}

A large amount of grains is a suitable instance of such a situation, amenable to be studied in the laboratory. When a dense flow of grains moves through an orifice, they are prone to clog due to the formation of an arch that blocks the exit \cite{To,Eric,Zuriguel,Aguirre,Longjas}. The properties of these arches have been characterized in previous works \cite{To2,Garci} and the main findings can be summarized in two points: 1) large arches tend to be semicircular in average, i.e. their span is twice their height. This implies that there exists a direct relationship between all the geometrical features (such as the number of beads, the span, and so on). 2) There is a considerable number of particles which are suspended from above the equator, stabilized by frictional forces. Concerning this point, in \cite{Garci} the angle associated to each grain, $\phi$, was defined as the one subtended by the two segments connecting the center of the sphere with the centers of its two neighbors (see Fig.~\ref{fig:sistexp}, \emph{inset}). It was reported that 17\% of the spheres hang from above the equator ($\phi>180^{\circ}$) and were called \textit{defects}. Of course, only friction can stabilize a grain in this way.

The clogging of silos due to arch formation is often avoided in
practical situations, such as industrial silos, by imparting a
vibration to the container \cite{Janda,Mankoc,Langston,Valdes}. The vibration is likely to impose incompatible stresses on the arch and to break it. In this paper, we take a closer look into this procedure and address the question of whether there exists a relationship between the geometric characteristics of the arches and the external vibration needed to break them. This can obviously lead to an improved efficiency in the above mentioned industrial processes, and also to understand
better which are the main ingredients affecting the stability of
arches. These results could also be related with several situations,
such as compaction dynamics \cite{Lumay,Pugnaloni3}, where the role of arches is crucial. For the moment, in this first approach, we have focused on the force needed to break an arch by increasing the
amplitude of the vibration, at a constant frequency.

We have set up an experimental device consisting of a two-dimensional symmetrical silo placed on top of a magnetic shaker (Fig.~\ref{fig:sistexp}). The silo is made of two sheets of transparent polycarbonate (390 mm high and 80 mm wide) lined with a conductive coating to avoid electrostatic charges. The gap between the two sheets is of 1.2 mm, and we filled it with non-magnetic stainless steel beads of 1 mm diameter (in some runs we used brass beads of the same size). At the middle of the silo there is a horizontal partition with an orifice of 4.45 mm, dividing the container in two equal compartments. An electric motor can rotate the silo half a turn around the horizontal axis, through a junction allowing free vertical motion of the container. Additionally, a standard video camera continuously records the region of the outlet. A computer controls all the components with the following protocol. Starting from a situation where all the beads are in the bottom compartment, the silo is rotated half a turn around the horizontal axis, so that the beads start to fall through the orifice. The eventual formation of an arch that stops the flow is automatically detected by image analysis, and a photograph of the arch is taken and stored. Then, a sinusoidal vibration of 1 kHz frequency is switched on, and an amplitude ramp of approximately $0.09\;g/s$ ($g$ is the acceleration of gravity) is applied to the silo. This frequency was chosen because it is an order of magnitude bigger than the characteristic time that it takes for a bead to fall its own diameter from rest under the action of gravity. Typical amplitudes obtained are below the micron range, which is about the size of the asperities of the beads. We have checked that the residual transversal acceleration is always well below 10 \% of the vertical acceleration. The breaking of the arch is detected from the video signal, at which moment the maximum acceleration of the sinusoidal forcing $\Gamma$ is calculated from the calibrated input signal. The vibration is kept until all the beads are in the bottom half of the silo, and the procedure restarts.

The photographs of the arches (Fig.~\ref{fig:sistexp}, \emph{inset}) can be analyzed to obtain $\phi$ (the angle associated to each bead). The two beads at the end of the arch have not an associated angle and indeed, they are not considered to belong to the arch. From the particle positions one can also obtain the geometrical features of the arch (height, span, number of beads, and so on). Our first aim, then, was to try to establish a relationship between the shape of the arches (or any other geometrical characteristic) and the force needed to break them, to ascertain whether or not some kind of arches withstand the external vibration better than others. Our experiment is tailored to this aim, because it provides both the acceleration at the breaking point and a photograph of the arch.

\begin{figure}
\includegraphics[width=\columnwidth]{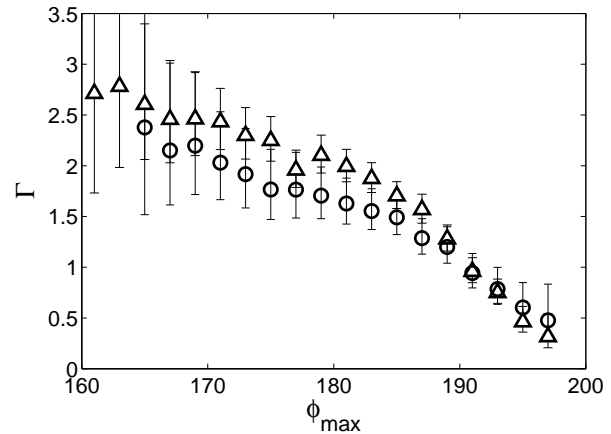}
\caption{The average maximum acceleration $\Gamma$ imposed by the
external vibration at the instant of arch breaking, as a function
of $\phi_{max}$, the maximum angle in the arch. Triangles correspond
to steel beads, and circles to brass beads. Error bars are 95\%
confidence intervals.} \label{fig:gamma}
\end{figure}

The first noticeable result is that the acceleration needed to break an arch, $\Gamma$, decreases with $\phi_{max}$, which is the maximum angle found among the beads belonging to a given arch (Fig.~\ref{fig:gamma}). To further substantiate the claim that the grain with $\phi_{max}$ is the weakest link in the arch, and that it sets the value of $\Gamma$ that the arch can withstand, we have inspected two hundred high-speed recordings of the breaking process (from the arch formation to the arch breaking). We have observed that 64\% of the arches break at the bead with maximum angle; 12\% break at a bead touching the one with $\phi_{max}$; another 12\% break at the border (where the value of $\phi$ cannot be defined) and the rest break elsewhere. More significantly, we observe that if an arch has a defect, in 95\% of the cases it breaks just there \cite{suppl}. It is therefore quite natural that as the arch breaks at the weakest link, \emph{i. e.} the grain with the largest $\phi$, then the value of $\Gamma$ depends on $\phi_{max}$ for each arch.

\begin{figure}
\includegraphics[width=\columnwidth]{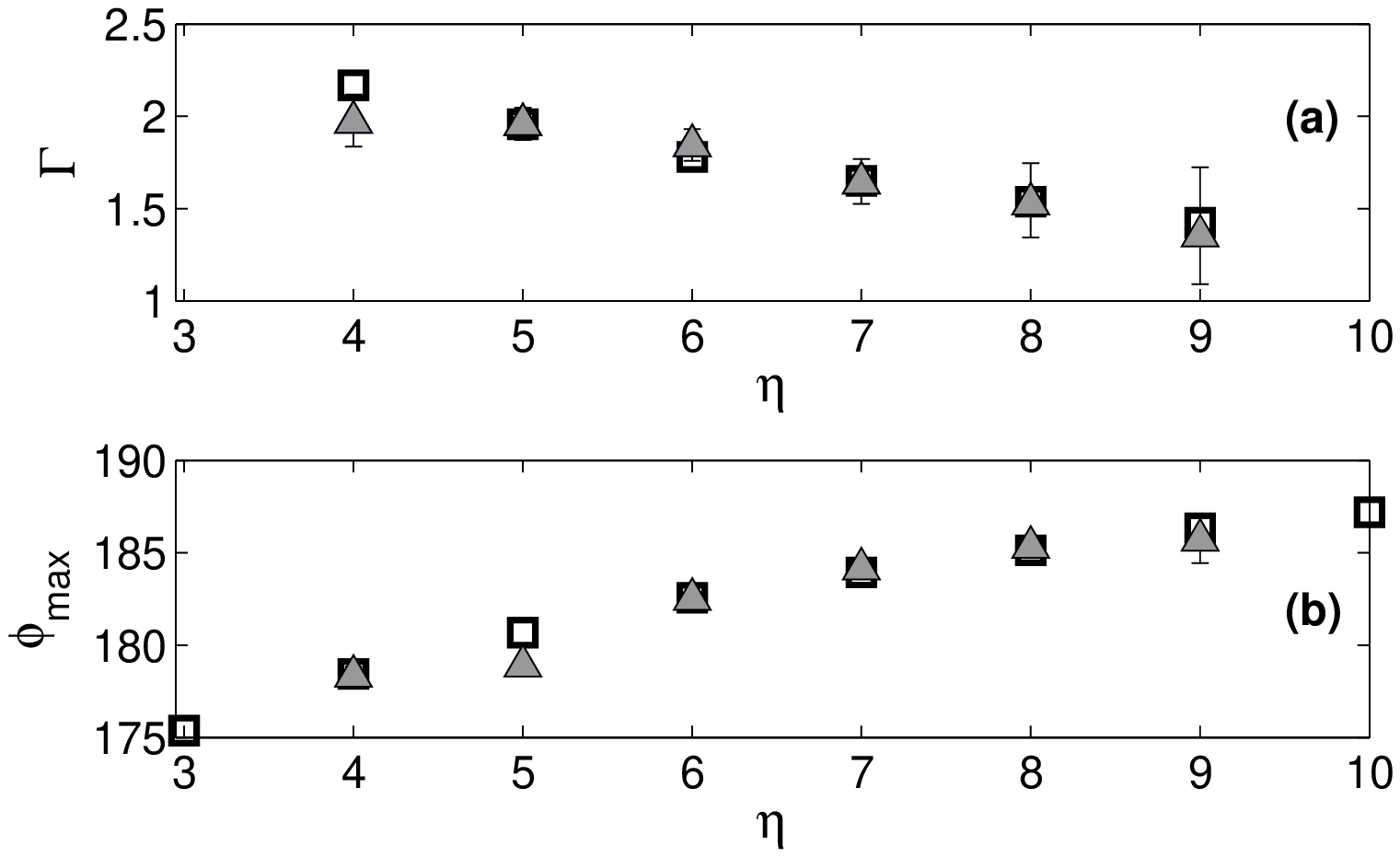}
\includegraphics[width=\columnwidth]{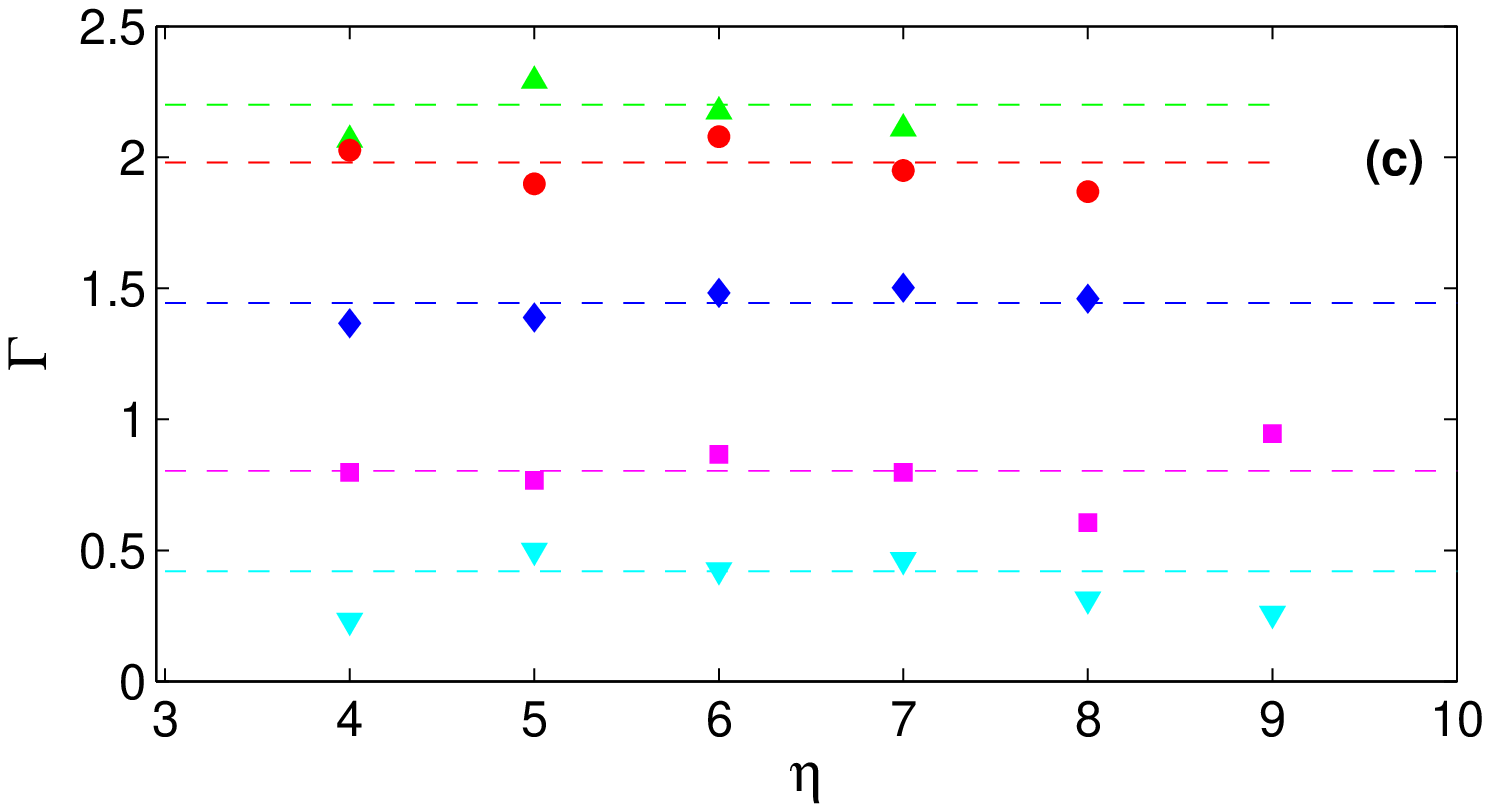}
\caption{(Color online)
\textbf{(a)} Value of $\Gamma$ versus the number of beads in the arch obtained experimentally ($\triangle$), and using the order statistics as explained in the text ($\square$).
\textbf{(b)} Average value of $\phi_{max}$ obtained for arches with different number of beads $\eta$ ($\triangle$). Expected value of $\phi_{max}$ ($\square$) calculated from the PDF of $\phi$, as a function of the sample size (the number of beads $\eta$).
\textbf{(c)}  $\Gamma$ vs. $\eta$ for small intervals of $\phi_{max}$. $\triangle: [173^{\circ},175^{\circ}]$;
$\circ: [179^{\circ},181^{\circ}]$;
$\diamond: [187^{\circ},189^{\circ}]$;
$\square: [191^{\circ},193^{\circ}]$;
$\triangledown: [195^{\circ},197^{\circ}]$.
Dashed lines correspond to the mean of the intervals. Data in this figure correspond to steel beads.
}
\label{fig:laprueba}
\end{figure}

\begin{figure}
\includegraphics[width=0.7\columnwidth]{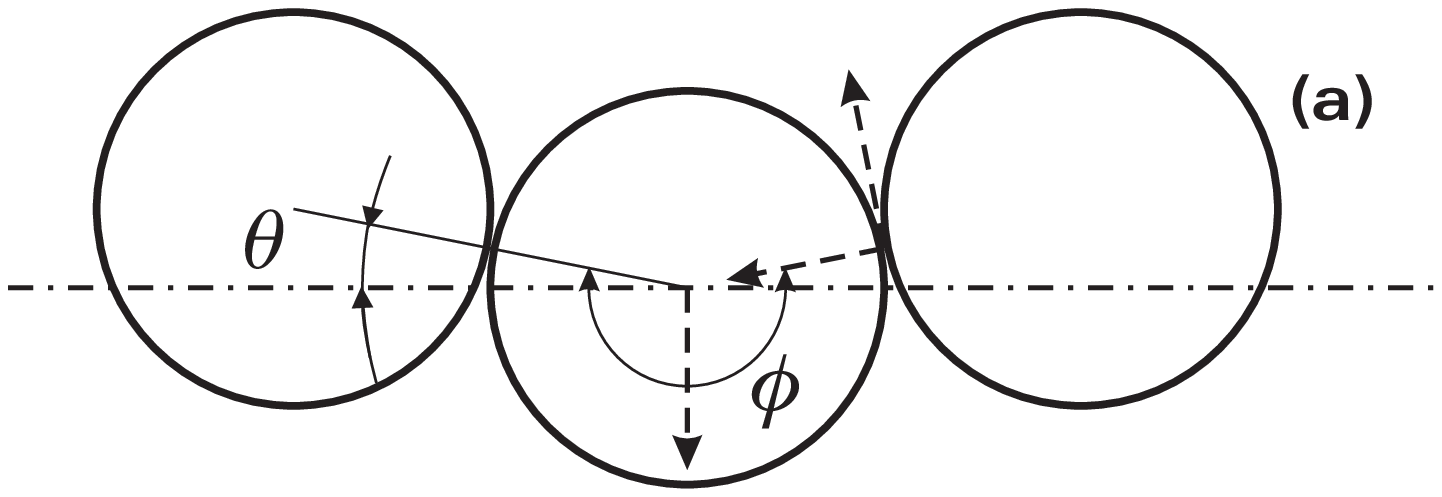}
\includegraphics[width=0.9\columnwidth]{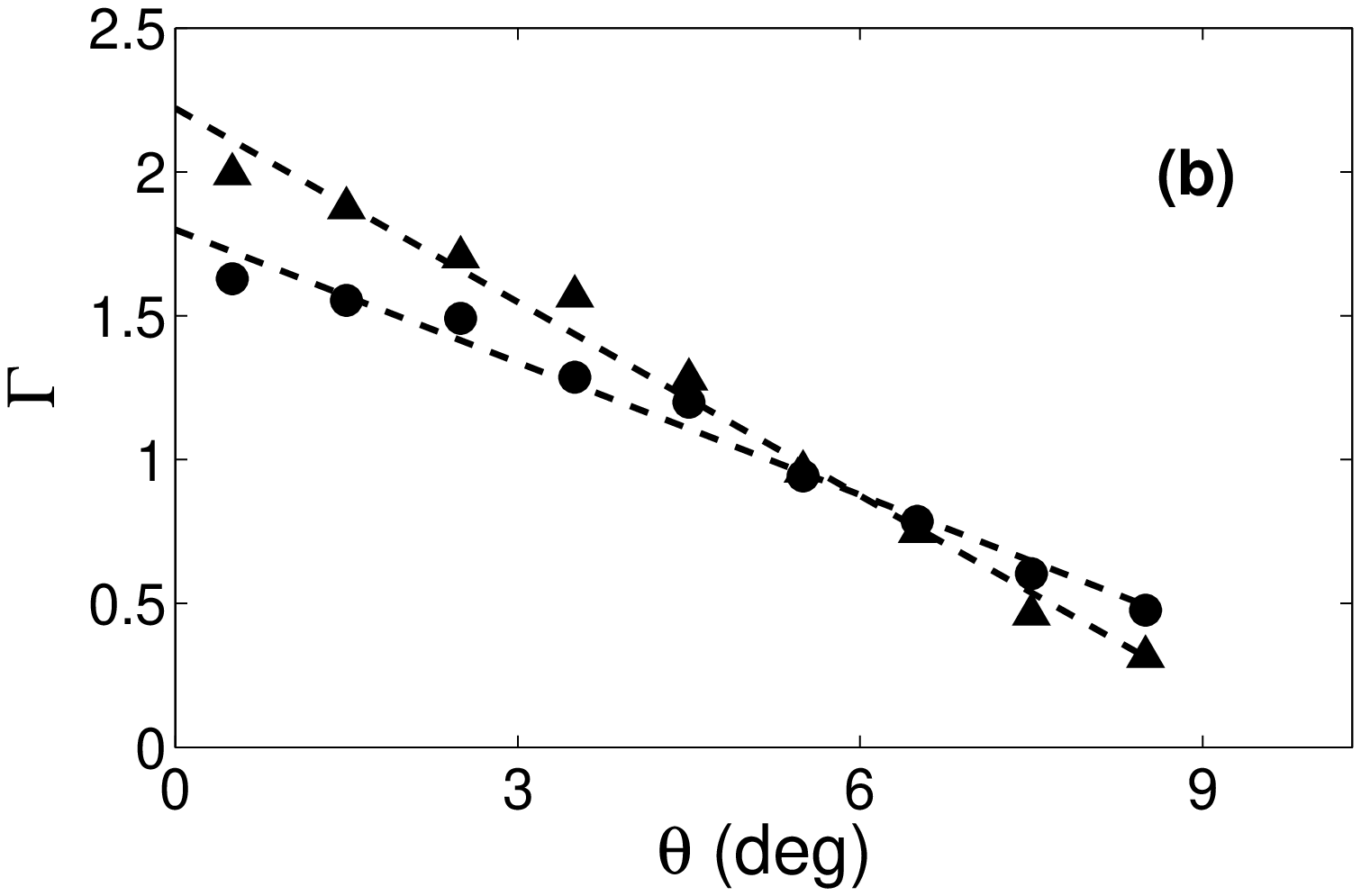}
\includegraphics[width=0.9\columnwidth]{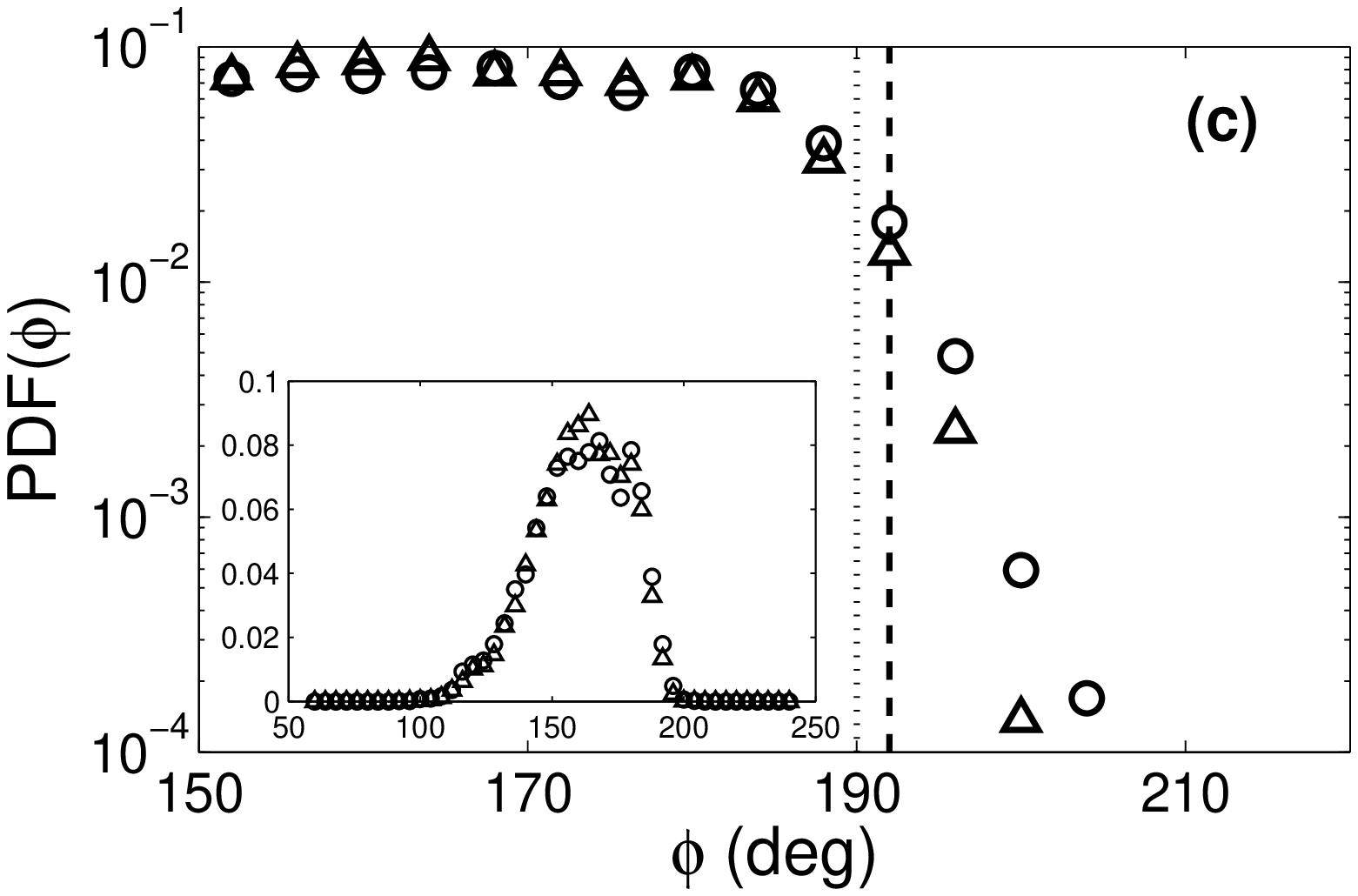}
\caption{\textbf{(a)} Diagram of a defect as considered in the
text. Forces for the right side of the bead are represented with
dashed vectors, from left to right: weight, normal force and
friction. \textbf{(b)} $\Gamma$ as a function of $\theta$ (same
data that in Fig.~\ref{fig:gamma}; triangles correspond to steel,
and circles to brass) along with least-squares fits (\emph{dashed
lines}). \textbf{(c)} Histogram for the angles $\phi$ of steel
beads ($\triangle$) and brass ($\circ$), in logarithmic scale. The
dotted vertical line corresponds to $\phi_c=190^{\circ}$ (steel) and the dashed one to $\phi_c=192^{\circ}$ (brass). In the inset, the whole histogram is shown in linear scale.}
\label{fig:rectas}
\end{figure}

Once the dependence of $\Gamma$ on the maximum angle found in the arch has been revealed, one could examine whether $\Gamma$ also depends on other properties of the arch (span, number of beads, etc.). Recall that all the geometrical features of the arches are directly related among them. Therefore we can focus on one of them, for instance, the number of beads $\eta$ (excluding, as remarked above, the two end grains). In Fig.~\ref{fig:laprueba}(a) we plot $\Gamma$ vs. $\eta$ for steel beads. Clearly, the more beads in the arch, the easier it is to break it. But this is just because as $\eta$ increases, the more likely it is that a larger $\phi$ appears. To illustrate this we have calculated the order statistics of $\phi_{max}$, \emph{i. e.} the expected value for the maximum of $\phi$ among a set of size $\eta$, as calculated from the PDF (probability distribution function) of $\phi$ (Fig.~\ref{fig:rectas} (c), \emph{inset}). These values are plotted along with the measured ones (see Fig.~\ref{fig:laprueba}(b)). As one can see, the agreement is good, meaning that the factor by which $\phi_{max}$ grows as beads are added to the arch can be understood just in statistical terms.

This implies that the angles of the arch are a random sample of the PDF of $\phi$. If the expected value of $\phi_{max}$ is used to estimate the average acceleration needed for breaking the arch (from Fig.~\ref{fig:gamma}) one obtains a prediction of $\Gamma$ as a function of $\eta$ that agrees quite well with the experimental results (Fig.~\ref{fig:laprueba} (a)). Besides, one can select the arches within a small interval of $\phi_{max}$ (although this reduces drastically the number of samples
considered) and see whether $\Gamma$ depends on $\eta$ in this subsample. It is hard to ascertain any dependence, as can be seen in Fig.~\ref{fig:laprueba} (c). Hence we can conclude that $\Gamma$ depends on the geometrical features of the arch only to the extent that these variables enhance or reduce the probability of finding a larger $\phi_{max}$.

In what follows, we perform an analysis of the forces in one bead that can explain the dependence of $\Gamma$ on $\phi_{max}$ for the case of defects (\emph{i.e.} beads hanging at $\phi>\pi$). Let us consider a bead hanging from above the equator from two neighbors in a horizontal, symmetric arrangement (see Fig.~\ref{fig:rectas} (a)). Certainly this is not always the case, and therefore this explanation can only be deemed as approximative. Considering the beads depicted in Fig.~\ref{fig:rectas} (a), the normal force $N$ can be shown to be almost equal (in first approximation) at both sides of the bead, and we can work out the force balance for half a bead when the arch breaks. At this moment, the friction is mobilized and there is an external force due to the vibration $F = m \Gamma g $, where $m$ is the mass of one bead. Let us define $\theta=(\phi-\pi)/2$ (see Fig.~\ref{fig:rectas} (a)). Therefore

\begin{equation}
\mu N \cos(\theta) = N \sin(\theta) + \frac{m g}{2} + \frac{m \Gamma g}{2}.
\end{equation}

Here $\mu$ is the friction coefficient. Assuming that $\theta$ is small, which in fact it is, we can write

\begin{equation}
\Gamma \approx - \frac{2N}{m g} \theta + \frac{2N}{m g} \mu -1.
\label{eq:gamma}
\end{equation}

From this simple model we can infer that there should be a linear
relationship between $\Gamma$ and $\theta$ (or $\phi_{max}$, because it depends linearly on $\theta$). In Fig.~\ref{fig:rectas} (b) we show that this linear relationship is fulfilled. From these data, corresponding to two different materials, we obtain $N=6.1 \pm 0.5\;mg$ and $\mu=0.26\pm0.08$ for steel, and $N=4.1\pm0.5\;mg$ and $\mu=0.35\pm0.08$ for brass. Unfortunately, it is not easy to provide accurately the value of static friction coefficient of the materials used: values given usually vary grossly depending on surface condition, lubrication, particle shape or sphericity, temperature, and so on (as we have observed in this experiment). Widely used values \cite{CRC} are compatible with the obtained ones. In any case, the relevant fact is that the proposed argument reproduces the dependence of $\Gamma$ on $\phi$ and yields reasonable values for $\mu$.

Even more momentous than the values of the friction coefficient is
the prediction that the normal force in a defect should be about
just a few times the weight of one bead. One could expect that the average pressure at the bottom of a silo, even taking into account the Janssen effect, should be much bigger than this. Indeed in a recent paper \cite{Pugnaloni2} the pressure is proposed to be much bigger than the value we give for the normal force in a defect. Hence, our result suggests that individual particles forming a defect are submitted to normal forces smaller than the average. It would we worth checking our
prediction either in experiments (which could be quite involved) or in numerical simulations.

Finally, if we set $\Gamma=0$ in Eq.~(\ref{eq:gamma}), we obtain
\begin{equation}
\phi_c=\pi+2\mu-\frac{mg}{N},
\end{equation}
which gives $\phi_c=192^{\circ}$ for brass and $\phi_c=190^{\circ}$ for steel. These angles correspond to the stability threshold in the limit of infinitesimal vibrations. In Fig.~\ref{fig:rectas} (c) the right side of the $\phi$ PDF is presented. A clear cutoff is obtained around the predicted values setting $\Gamma=0$ in Eq.~(\ref{eq:gamma}). This reinforces the validity of the explanation offered.

We have presented experimental evidence showing that arches break
at the weakest link, which is the bead in the arch which clings to its neighbors with the highest angle. Indeed, the value of this maximum angle in the arch is the best predictor for the force needed to break it when submitted to an external vibration. We provided a simple
argument to show that a direct relationship is expected between
the acceleration $\Gamma$ and the maximum angle $\phi_{max}$, in
the case $\phi_{max}>180^{\circ}$. These results open an array of
various questions, such as the study of the influence of the
vibration frequency on the arch stability, or the expected smaller
time lapse that a big defect would endure before the arch breaks.
Besides, very hard arches should be obtained when defects are absent.

\begin{acknowledgments}
We thank D. Maza, R. Cruz and L. A. Pugnaloni for their comments and discussions, and L.F. Urrea for technical help. This work has been financially supported by Projects FIS2008-06034-C02-01 and FIS2011-26675 (Spanish Government), and PIUNA (Universidad de Navarra). G.L. thanks the F.R.S-FNRS for the financial support and C.L. thanks Asociaci\'on de Amigos de la Universidad de Navarra for a scholarship.
\end{acknowledgments}

\end{document}